\documentclass{emulateapj}

\usepackage{natbib}
\usepackage{graphicx,amssymb}

\usepackage[dvipsnames]{color}

\shorttitle{The roles of radiation and ram pressure in driving galactic winds}
\shortauthors{Sharma, Nath}

\begin{document}

\newcommand{\half}{\frac{1}{2}}
\newcommand{\3}{\ss}
\newcommand{\n}{\noindent}
\newcommand{\eps}{\varepsilon}
\def\be{\begin{equation}}
\def\ee{\end{equation}}
\def\ba{\begin{eqnarray}}
\def\ea{\end{eqnarray}}
\def\de{\partial}
\def\msun{M_\odot}
\def\div{\nabla\cdot}
\def\grad{\nabla}
\def\rot{\nabla\times}
\def\ltsima{$\; \buildrel < \over \sim \;$}
\def\simlt{\lower.5ex\hbox{\ltsima}}
\def\gtsima{$\; \buildrel > \over \sim \;$}
\def\simgt{\lower.5ex\hbox{\gtsima}}
\def\etal{{et al.\ }}
\def\red{\textcolor{red}} 
\def\blue{\textcolor{blue}}
\def\del{\partial}
\newcommand{\pd}{\partial}


\title{The roles of radiation and ram pressure in driving galactic winds
}

\author{Mahavir Sharma, Biman B. Nath}
\affil{Raman Research Institute, Sadashiva Nagar, Bangalore 560080, India}
\email{mahavir@rri.res.in; biman@rri.res.in}

\begin{abstract}
We study gaseous outflows from disk galaxies driven by the combined effects of ram pressure on cold gas clouds and 
radiation pressure on dust grains. Taking into account the gravity due to disk, bulge and dark matter halo, and assuming
continuous star formation in the disk, we show that radiation or ram pressure alone is not sufficient to drive escaping winds 
from disk galaxies, and that both processes contribute.  
We show that in the parameter space of star formation rate
(SFR) and rotation speed of galaxies, the wind speed in galaxies with rotation speed $v_c\le 200$ km s$^{-1}$  and
  SFR $\le 100$ M$_{\odot}$ 
yr$^{-1}$, has a larger contribution from ram pressure, and that in high mass galaxies 
with large SFR, radiation from the disk has a greater role in driving galactic winds. The ratio of wind speed to circular speed can
be approximated as
${v_w \over v_c} \sim 10^{0.7} \, \left[{\rm SFR\over 50 \, {\rm M}_{\odot} \, {\rm yr}^{-1}}\right] ^{0.4} \ \left[{v_c\over 120\, km/s}\right]^
{-1.25}$.
We show that this conclusion is borne out by 
observations of galactic winds at low and high redshift and also of circumgalactic gas.
We also estimate the mass loading factors under the combined effect of 
ram and radiation pressure, and show that the ratio of mass loss rate to SFR scales
roughly as $v_c^{-1} \Sigma_g^{-1}$, where $\Sigma_g$ is the gas column density in the disk.
\end{abstract}


\section{Introduction}
Galactic winds have been observed at different wavelengths in galaxies of various masses and in a range of redshifts. Galaxies, especially with star formation rates ($\Sigma_{SFR}$) $\ge 10^{-1}$ M$_{\odot}$ yr$^{-1}$ kpc$^{-2}$, often show large outflow of hot gas that emits X-rays and in which cold clouds are found to be embedded, which are observed with H$\alpha$ or NaD  lines \citep{hec00,mar05}. The speed of the clouds in the wind range from a few tens to several hundred km s$^{-1}$, and the total mass loss rate can be several times the star formation rate \citep{vei05}.

These outflows play a crucial role in the evolution of galaxies by expunging gas,
and thereby suppressing the star formation. The attempts to understand galactic
evolution in the cosmological context have since long encountered the so-called
'cooling catastrophe' problem, since left to its own device the baryonic gas would cool and form
stars more rapidly than observed. It is generally believed that a feedback loop inhibits this, and that the process of star formation excites an outflow and quenches itself.
The observed mass-metallicity relation in galaxies also indicate that galactic outflows play a major role in
the chemical evolution in galaxies. Furthermore, these outflows enrich the intergalactic medium with 
metals.

The standard model to understand galactic outflows involves a heated interstellar
medium (ISM) under the influence of supernovae (SN), and the hot gas  being driven by thermal pressure \citep{che85, hec02}. The expansion speed of
this hot gas can be large enough to eject it out of  the galaxy \citep{lar74,sai79,dek86}. 
The observations of cold gas in these outflows \citep{hec00} led to the proposal
that the cold gas entrained in the hot gas moved due to ram pressure. The wind speed was however not found to correlate with galaxy mass \citep{hec00,mar99}, and it was argued that the supernovae rate increased with SFR and hence the wind velocity might correlate with SFR.  Simulations  also supported this scenario \citep{suc94,str00}. However, there is a limiting cloud speed implicit in this
process since ram pressure acts on the cold gas until the cold gas velocity becomes equal to that of hot gas.

This scenario, however, has met with problems from new observations of cold component which show that the terminal outflow speed  depends on galactic properties like rotation speed \citep{mar05,rup05}.
It has been proposed that these observations can be explained  by radiation pressure driving the outflow  
\cite{mqt05, mar05, sha11}.
It has also been pointed out that a natural course of events leading from a starburst would be a radiation pressure driven wind in the beginning, and ram pressure acting on it after a period of $\sim 3\hbox{--}5$ Myr, the life time of massive stars \citep{nath08, mur11}. This scenario also naturally explains the puzzling fact that cold clouds are observed at large distances although their survival time-scales in the hot gas would have inhibited them from being pushed out to such distances.

In the face of two processes leading to outflows, one wonders if both processes contribute equally, or if there are regimes in which one of these two processes dominate over the other. 
In this paper we present an analytical calculation for the dynamics of cold clouds taking into
account both ram and radiation pressure and all sources of gravity, and compare our results with
observations.

\section{Gaseous outflows with ram and radiation pressure}
We consider the dynamics of cold clouds ($T \lesssim 10^4$ K)
embedded in hot gas, 
in which the hot gas component exerts a drag force due to ram pressure. We also
assume that dust grains in the cold clouds are strongly coupled to the gas,
and therefore the dynamics of these clouds is also influenced by radiation pressure.
We ignore magnetic forces and the compression of cold clouds by hot wind gas.

\begin{figure}[h]
\includegraphics[angle=0,scale=.50]{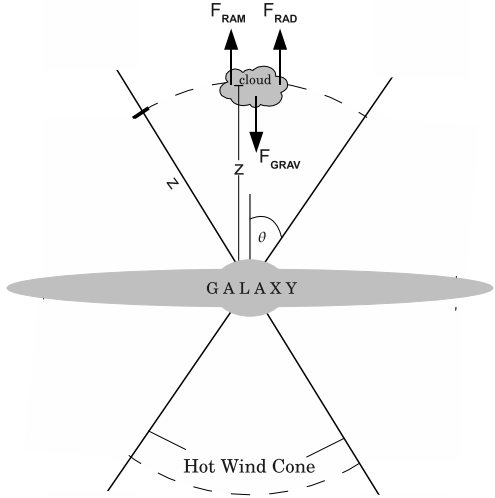}  
\caption{A schematic diagram for the motion of a cold cloud embedded in a cone of hot flow and acted
upon by forces of radiation and gravity from the parent galaxy. Cloud is at a height z. The total mass loss in hot flow, $\dot M_h$ = 2 $\rho_h v_h$A, where A is the area at the top of conical patch, and a factor of 2 for two-sided mass loss. For a half-cone angle of $\theta$ the area  A = z$^2$ $\int_0^{2\pi}\int_{0}^{\theta}$sin($\theta '$)d$\theta ' d\phi$.  
         }
\label{cloud}
\end{figure}
We therefore solve the following equation for the momentum of the cold cloud (P$_c$ = M$_c$v), see  figure \ref{cloud},

\be
{dP_c \over dt } = M_c \frac{d v}{d t} = F_{ram} + F_{rad} - F_{grav}
\label{one}
\ee
where M$_c$ is the mass of the cloud and v is its velocity  in z direction. F$_{ram}$ represents the force exerted by the hot wind via ram pressure in g cm s$^{-2}$. F$_{rad}$ is the force due to radiation on dust grains and F$_{grav}$ is the gravitational force.

We first discuss the role of ram pressure on the motion of cold blobs of
gas dragged in it, following the model of
\cite{str73}.
In this scenario, the hot gas observed in
X-rays and which is thought to provide the ram pressure,
has temperatures in the range
$0.5\hbox{--}1$ keV \citep{hec02,mar05}, which correspond to the
 isothermal sound speed
$c_s \sim 300\hbox{--}400$ km/s.
Current X-ray instruments can not detect the speed of this hot and tenuous
material and hence the kinemetics of  this hot phase is poorly constrained.
If  we assume it as an adiabatic wind passing through a sonic point, then
$v_h^2
\sim \alpha c_s^2$, where $\alpha = 2.5 \hbox{--}5$ \citep{efs00}, which gives
$v_h\sim1.2\hbox{--}2.2 \, c_s$. In this
paper, we take
v$_h \sim 800$ km s$^{-1}$, which corresponds to $v_h \sim 2 c_s$ and $T_X \sim
1$ keV.

Consider then
the hot gas flow (with density $\rho_h$ and velocity $v_h$), emerging through a cone. Mass loss in a hot wind is given by the following expression (see figure \ref{cloud}),
\begin{equation}
 \dot M_h = 2 \rho_h v_h z^2 \int_0^{2\pi}\int_{0}^{\theta}sin(\theta ')d\theta ' d\phi
\end{equation}
Observations 
indicate conical angles for hot wind  in the range $2\theta \sim 10^\circ\hbox{--}100^\circ$ \citep{vei05,leh96}. We consider a mass loss rate of $\dot M_h\approx  (\pi/2) z^2\rho_h v_h$, which roughly corresponds to half-cone angle $\theta \sim 30^\circ$.
The momentum injection rate is  $\dot p_h = \dot M_h v_h$, so we can write
\begin{equation}
 \rho_h v_h^2 =\frac{\dot p_h}{\pi z^2 /2}
\label{IMDI}
\end{equation}
The force exerted by the ram pressure on a
cold cloud of mass M$_c$ and cross-section A$_c$ is given by
\ba
F_{ram} &&= \half C_D A_c \rho_h (v_h - v)^2\, \mathcal{H}(v_h-v)
\nonumber \\
&&=\frac{C_D A_c}{2} \rho_h v_h^2  \left(1 - {v\over v_h}\right)^2\, \mathcal{H}(v_h-v)
\label{FRAM}
\ea
Here $\mathcal{H}(v_h-v)$ is the  step function whose value is 1 for $v < v_h$ and 0 otherwise. $C_D \sim 0.5$ is the drag coefficient.
For the cloud, one can write $\frac{M_c}{A_c} = \mu m_p N_H$
 where N$_H$ is the column density and $\mu$ is the mean molecular weight.
 Also the momentum injection rate $\dot p_h$   is 
$\sim \left[5 \times 10^{33}\, \left({{\rm SFR} \over {\rm 1\, M}_\odot / {\rm yr}}\right)\right]{\rm dyne}$  
in a starburst \citep{leith99}.  Using these and substituting eqn.(\ref{IMDI}) in eqn.(\ref{FRAM}) we get,
\be
\frac{F_{ram}}{M_c} = \frac{[5\times10^{33} \,  ({SFR \over 1\, M_\odot/yr})]{\rm dyne}}{4\  N_H\ \mu m_p \  (\pi z^2/2)}\left(1-\frac{v}{v_{h}}\right)^2 \, \mathcal{H}(v_h-v)
\ee

Next we consider the forces due to a galactic disk. We will use $f$ for force per unit mass ($f={F \over M_c}$).
In cylindrical geometry, the force
of gravitation $f_{g,d}(z)$, and that due to radiation $f_{r,d} (z)$, along the pole of a disk of radius r$_d$, with
constant surface density ($\Sigma$) and surface brightness ($I$) 
 are given by,
\be
f_{g,d}=2 \pi G \Sigma\int^{r_d}\frac{zrdr}{(r^{2}+z^{2})^{3/2}} \,,
f_{r,d}=\frac{2 \pi\kappa I}{c}\int^{r_d}\frac{z^{2}rdr}{(r^{2}+z^{2})^{2}} \,,
\label{eq:fg}
\ee 
where  $\kappa$ is the average opacity of a dust \& gas mixture.
The ratio of these forces,
the Eddington ratio, increases with the height $z$, beginning with a value
of $\Gamma_0 = {\kappa I \over 2 c G \Sigma}$ at the disk centre at $z=0$.
Since
${I\over \Sigma} \propto {L\over M_d}$, where M$_d$ is the disk mass, we can
express $\Gamma_0$ in terms of the SFR by
 calculating the luminosity $L$ of a galaxy
in any desired band for a certain SFR using the Starburst99 code. 
The luminosity in this case is proportional to SFR, therefore if 
$L_1$ is the luminosity at $1$ Gyr for an SFR of $1$ M$_\odot$ yr$^{-1}$ 
then we can write $\Gamma_0$ as,
\be
\Gamma_0 = {\kappa \over 2 c G}{L_1 \times {SFR\over 1 M_\odot/yr} \over M_d}
\label{eq:gamma1}
\ee  
 We use the mean opacity for
gas mixed with dust $\sim 200$ cm$^2$ g$^{-1}$ corresponding to a color temperature  $\sim 9000$ K in the U band (Figure 1b, \cite{dra11}.

\begin{figure}[h]
\includegraphics[angle=0,scale=.40]{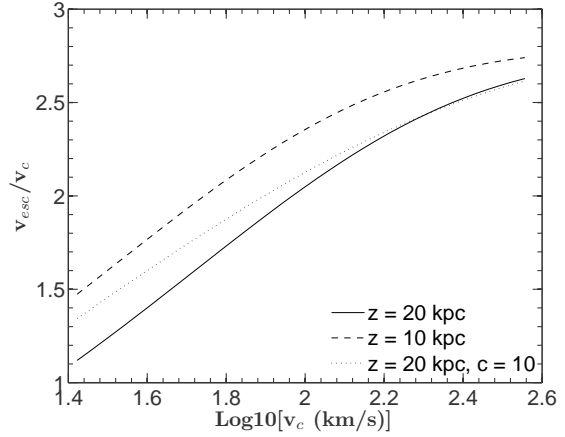}  
\caption{Ratio of NFW escape speed to the circular speed vs.  the circular speed, at two different vertical distances, $10$ kpc (dashed line), and $20$ kpc (solid line). Dotted line is for a fixed value of halo concentration parameter 
$c=10$.
         }
\label{figesc}
\end{figure}

To determine the gravitational force, we assume a spherical mass 
distribution in the bulge and
halo. For the bulge, we assume a total mass of $M_b \sim 0.1 M_d$ inside a radius
$r_b \sim 0.1 r_d$ for simplicity.  
For the halo, we consider a Navarro-Frenk-White (NFW) profile, 
with total mass $M_{vir}$ \citep{nfw97}.
We fix M$_{vir}$ for a given disk
mass ($M_d$), by the ratio $M_{vir}/M_d\sim 20$, as determined by
Mo, Mao \& White (1998) (referred to as MMW98 hereafter). 
We evaluate the disk exponential scale-length (r$_d$) using the prescription of MMW98,
and use it as the size of galactic disk.
Gravitational potential of  NFW halo is,   
\be
\Phi_{NFW} = -\frac{G M_{vir}}{\ln(1+c)-c/(1+c)} \left[\frac{\ln{(1+\frac{R}{R_s})}}{R}\right]
\ee
where 
R = $\sqrt{r^2+z^2}$,
$c=\frac{R_{200}}{R_s}$ is the concentration parameter,  R$_s$ is the NFW scale length and R$_{200}$ is the radius within which the mean overdensity is 200.
This potential implies a gravitational force along z which is given by, 
\begin{eqnarray}
f_{halo,z}&=&\left|-\frac{\pd \Phi_{NFW}}{\pd z}\right|_{r=0} \\ \nonumber 
&=& {G M_{vir} \over z^2} \left[{\ln(1+{z\over R_s})-z/(z+R_s) \over \ln(1+c) - c/(1+c)}\right]
\end{eqnarray}
The rotation speed implied by the NFW profile peaks at a radius $R\sim 2 R_s$, given by,
\be
v_c ^2=v_{200}^2 \,{c \over 2} \, {\ln (3)-2/3 \over \ln (1+c) -c/(1+c)} \,,
\ee
where  $v_{200}$ is the rotation speed at R$_{200}$.
We choose this value of the maximum rotation speed
to represent the $v_c$ of the disk galaxy, since Figure 2 of MMW98 shows that the value of $v_c$
from the flat part of the total rotation curve does not differ much from the peak of the
rotation curve from halo only. 
The escape speed in a NFW halo is given by 
\be
v_{esc}^2 = v_c^2 \left[{4 \over \ln(3) - {2 \over 3}}
\left({R_{200}\over R}\ln(1 +{c R\over R_{200}})-{c\over 1+c}\right)\right] \,.
\ee
Figure $\ref {figesc}$ shows the escape speed along the z-axis for different galaxies. The dashed and solid lines show the escape speed at $10$ and $20$
kpc from the disk plane, for galaxies with different circular speed. We have 
used the relation between the halo concentration parameter $c$ and galactic mass
as given by \cite{mac05}.
We find that for low mass galaxies with v$_c \leq 100$, the escape velocity v$_{esc} \lesssim 2 v_c$, and that for higher mass galaxies, the escape speed
ranges between $2\hbox{--}3 \, v_c$. We can therefore conclude that
for escaping winds, the ratio of wind speed
to circular speed should be in the range of $2\hbox{--}3$.

One can finally rewrite eqn(\ref{one}) for evaluating the velocity of clouds as a function of z, 
\begin{eqnarray}
v{d v \over d z}= && \frac{[5\times10^{33}\, (\frac{SFR}{1 M_\odot/yr})]{\rm dyne}}{4 N_H\ \mu m_p \  (\pi z^2/2)}\left(1-\frac{v}{v_{h}}\right)^2 \, \mathcal{H}(v_h-v)\, \nonumber \\
&&+ 2 \pi G \Sigma \Gamma_0\left({r_d^2 \over z^2 +r_d^2}\right) - 2\pi G\Sigma \left(1-{z \over \sqrt{z^2 +r_d^2}}\right) \, \nonumber \\
&&- {G M_b \over z^2} - {G M_{vir} \over z^2}\left({\ln(1+{z\over R_s})-{z\over z+R_s} \over \ln(1+c) - \frac{c}{1+c}}\right)
\label{eq:main}
\end{eqnarray}
where $\Gamma_0$ is given by eq. \ref{eq:gamma1}. We use $\mu=1.4$ and $N_H \sim 10^{21}$ cm$^{-2}$ \citep{mar05,hec00}.
Here the first term on RHS denote ram pressure, second  the radiation pressure and
the last three terms  represent 
the gravity of the disk, bulge \& NFW halo respectively.
This equation is  non-linear due to the presence of $v$ in ram pressure term 
and should be solved numerically, although previous authors have approximated
it assuming $v\ll v_h$.
The form of the ram pressure term suggests
that ram pressure would not be effective once the velocity becomes greater 
than velocity of hot component.
Hence the ram pressure is likely to be effective for low-mass galaxies.
\begin{figure}[h]
\includegraphics[scale=.45]{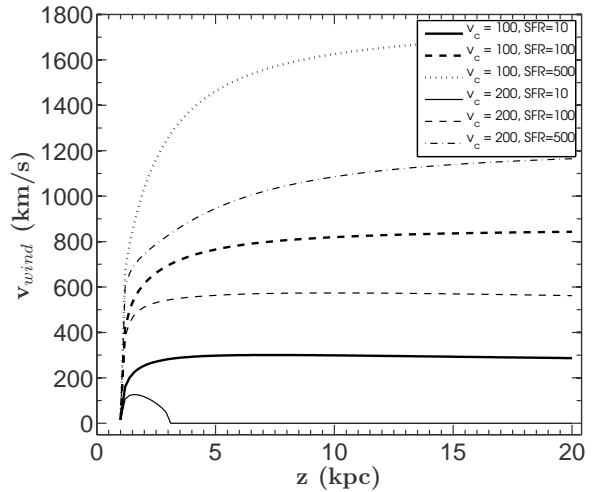}
\caption{
Variation of wind speed with vertical distance ($z$) for galaxies
of different circular speeds. The thick solid, dashed and dotted lines refer
to $v_c=100$ km s$^{-1}$, and for SFR of $10, 100, 500$ M$_{\odot}$ yr$^{-1}$,
respectively.
The thin solid, dashed and dot-dashed lines refer to $v_c=200$ km s$^{-1}$,
for the same values of SFR, respectively.
}
\label{figwzcomb}
\end{figure}

\section{Results}
We solve the wind equation (eqn \ref{eq:main}) numerically. Figure
$\ref{figwzcomb}$ shows the wind speed as a function of $z$ for different
values of SFR for two galaxies, with $v_c=100$ km s$^{-1}$ 
and $v_c=200$ km  s$^{-1}$. Instead of rising continuously,
the wind speed saturates after travelling a distance of $\ge
10$ kpc, with a terminal speed that is lower for higher mass galaxies. 
The thick solid line 
roughly corresponds to M82,
and the wind speed $\sim 300$ km s$^{-1}$ is consistent with 
observations \citep{hec00,sch04}.

\begin{figure}[h]
\includegraphics[angle=0,scale=.45]{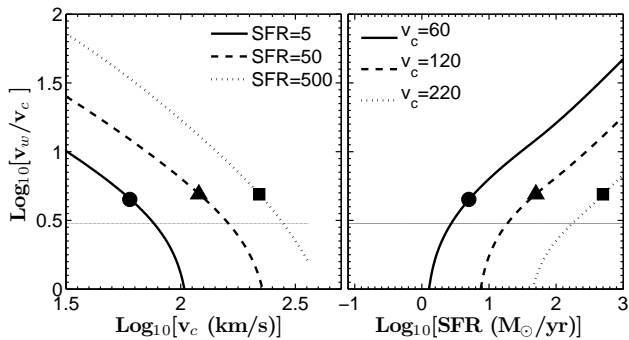}  
\caption{Ratio of wind velocity at 20 kiloparsec and the galactic rotation speed plotted 
with  v$_c$  for three different SFR in the left panel \&  with SFR for three different v$_c$ in the right panel.
Three representative cases are shown with solid circle (dwarf
starbursts), solid triangle (LIGs) and solid square (ULIGs). The thin horizontal line corresponds
to $v_w =3 v_c$.
}
\label{figcoef}
\end{figure}

We then use the wind speed at $z=20$ kpc and show the variation of $v_w/v_c$ with
circular speed $v_c$ and SFR
in the left and right panels of
Figure \ref{figcoef} respectively. We find that, for a constant SFR, $v_w/v_c$ decreases with $v_c$, as
gravity increases with $v_c$. We also show three representative cases in this plot,
of dwarf starbursts (solid circle:
$v_c\sim 60$ km s$^{-1}$, SFR $\sim 5$ M$_{\odot}$ yr$^{-1}$),
LIGs (solid triangle:
$v_c\sim 120$ km s$^{-1}$, SFR $\sim 50$ M$_{\odot}$ yr$^{-1}$),
ULIGs (solid square:
 $v_c\sim 220$ km s$^{-1}$, SFR $\sim 500$ M$_{\odot}$ yr$^{-1}$).
The values of $v_w/v_c$ lie close to $\sim 3$ which is shown by the thin horizontal
line. The near constancy of v$_w$/v$_c$ for the three representative points recovers the observed 
scaling of v$_w$ with v$_c$.
Taking into account the variation of v$_w$/v$_c$ with v$_c$ and SFR we find that, the results can be approximated by the following fit,
\begin{equation}
{v_w \over v_c} \sim  10^{0.7}\,
\left[{\rm SFR\over 50 \, {\rm M}_{\odot} \, {\rm yr}^{-1}}\right] ^{0.4} 
\left[{v_c\over 120\, km/s}\right]^{-1.25} \,.
\label{eq:fit}
\end{equation}
\begin{center}
 \begin{figure}[h]
\includegraphics[angle=0,scale=.45]{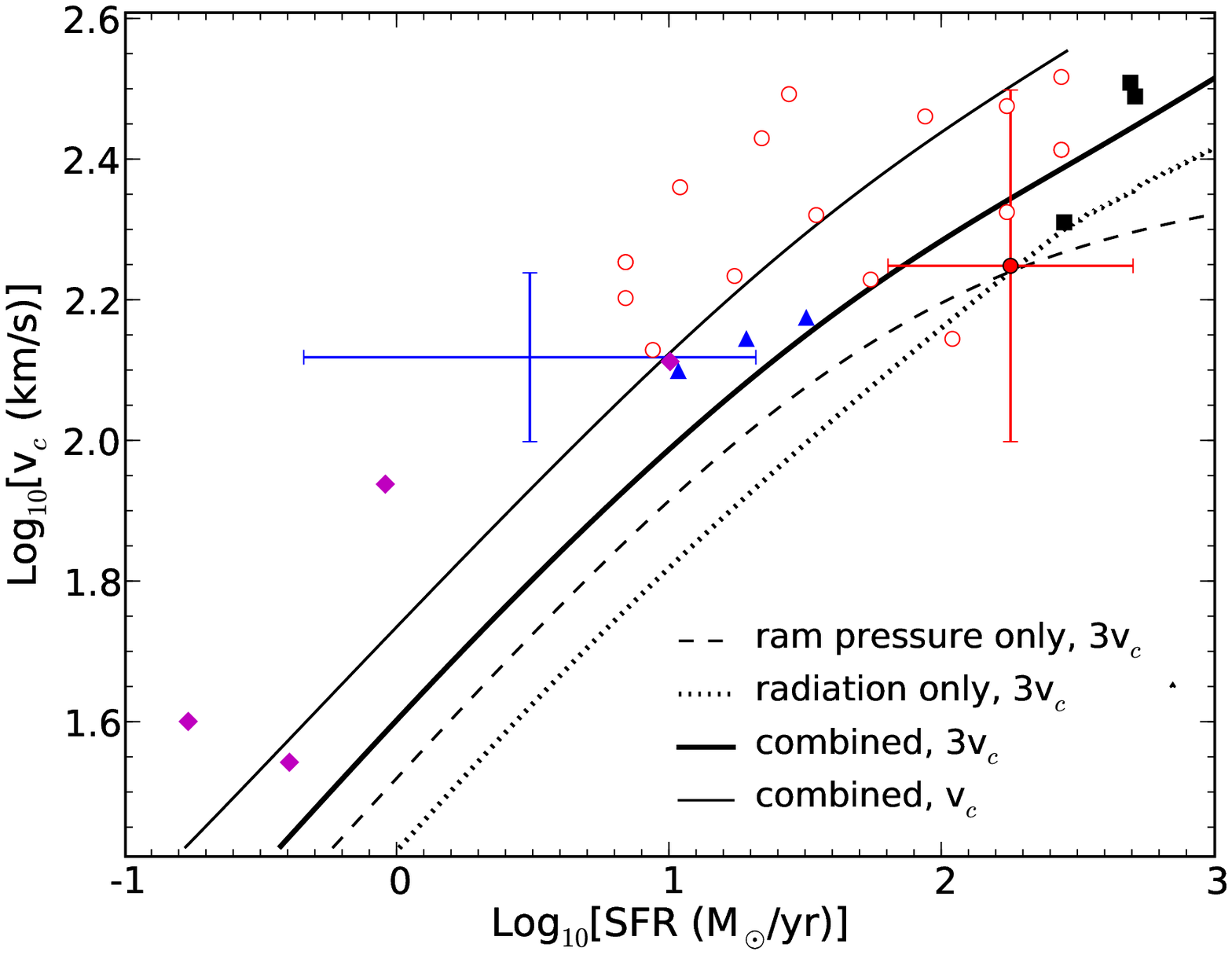}  
\caption{Contours of $v_{w,20 kpc}$ in units of v$_c$ for winds driven by only ram pressure, only 
radiation pressure \& the combination of two. The wind velocity increases as one moves from top left corner to bottom right corner.  Also plotted are the data points with different symbols: hollow circles \citep{hec00}, squares \citep{mar05, gen01}, 
triangles \citep{wei09}, diamonds \citep{sch04}, big red cross with circle at its center 
\citep{rup05}), big blue cross (\cite{tum11} without two outliers in SFR). 
        }
\label{figcie}
\end{figure}
\end{center}

Next we solve the wind equation for a grid of SFR and galaxy circular
speed values, for the cases of ram pressure and radiation pressure alone,
and then for the combination of the two. In Figure \ref{figcie} the wind velocity is zero in top left corner for high mass \& low SFR galaxies. Wind velocity increases as one moves diagonally, from top left to the bottom right corner.  We 
show two contours for $v_w=3 v_c$ with thin solid lines for
 ram and radiation pressure alone.
For the case of combined ram
and radiation pressure driving, we show two contours, for $v_w=v_c$ and $3v_c$
(upper and lower thick lines). We also show the data for outflows from
a number of observations (see caption for details).

In the case of only radiation pressure, the wind speed is found to be
roughly proportional to SFR, which can be understood from the fact that
$\Gamma_0 \propto {\rm SFR}$.
The case for only ram pressure appears to explain the wind in low mass galaxies. However, 
from the $v_w=3v_c$ contour it is clear that ram pressure can not drive the cold 
clouds out of the galaxies with rotation speeds $\gtrsim$ 200 km/s, as we have seen in
the previous section that for escaping winds one needs $v_w \sim 3 v_c$. 
This points to the existence of a critical rotation speed above which the physical mechanism of outflow changes. Therefore outflows from galaxies with v$_c \le 200$ km/s and SFR $\le 100$ M$_\odot$/yr are
 dominated by ram pressure and those from the more massive galaxies with larger SFR,  are influenced
more by radiation pressure. 

\section{Discussions}
The most important result of our calculation is that galactic outflows require
both ram and radiation pressure, especially for high mass and high SFR cases.
Our calculation has a number of ingredients from
stellar physics and disk and halo parameters, and apart from the value of the hot
wind speed $v_h$, there is no free parameter in this calculation. It is therefore
interesting to note that our theoretical results are consistent with most data
of outflows when studied in the parameter space of $v_c$ and SFR. It is also interesting
that a recent simulation with ram and radiation pressure driven outflows
has concluded that these two processes are important
in different mass regimes, 
although it is not clear where
the dividing line between the two regimes lies \citep{hopkins11,sch11}.
Cold cloud outflows from galaxies on the left of the contours in figure \ref{figcie} are unlikely to escape into the IGM and likely get trapped in the circumgalactic region as observed by \cite{tum11} (data shown by blue cross) or fall back \citep{opp08}. 

Although strictly speaking our calculation refers to cold clouds being driven out
along the pole of the disk galaxies, and we cannot infer the mass loss rate without 
doing a 2-D calculation, but
 we can speculate on the scaling
of the mass loss rate with galactic mass by making some simple 
assumption. Let us assume that the dynamics of cold clouds beyond the polar regions
are similar to that along the pole. Assuming a one-dimensional
mass flow, the mass loss rate from the disk is approximately $\dot{M}_w\propto v_w [\Sigma_g \pi r_d^2]$,
where $\Sigma_g$ is the gas column density  and $r_d$ is the scale length
of the disk. We note that in the prescription of MMW98, one has
$v_c \propto r_d$. We therefore have, $\dot{M}_w \propto v_c^{2-0.25} \dot{M}_\ast ^{0.4} \Sigma_g$,
where we have used eqn \ref{eq:fit}, after multiplying
both sides by $v_c$. The ratio
of mass outflow rate to the SFR is therefore $\dot{M}_w/\dot{M}_\ast \propto v_c^{1.75}
\Sigma_g \dot{M}_\ast^{-1.4}$. Using  Kennicutt's law of star formation,
which gives $\dot{M}_\ast \propto \Sigma_g^{1.4} r_d^2 \propto \Sigma_g ^{1.4} v_c^2$, we have finally,
$
{\dot{M}_w \over \dot{M}_\ast} \propto v_c^{1.05} \Sigma_g^{-0.96}
$.
We can therefore conclude that roughly,
\begin{equation}
{\dot{M}_w \over \dot{M}_\ast} \propto v_c^{-1} \Sigma_g ^{-1} \,.
\end{equation}
Interestingly, similar power law dependence has also been found in simulations \citep{hopkins11}.

We note that our results assumed a value of $v_h\sim 800$ km s$^{-1}$, and a column
density of cold clouds of $\sim 10^{21}$ cm$^{-2}$. If we assume
a larger value of $v_h$ ($\sim 1000$ km s$^{-1}$), then the contour for only ram pressure will be able to explain the winds in ULIGs
with large SFR and high mass. A similar result will follow from larger values of $\kappa$ for the radiation pressure case. 

It is interesting to note that the contour for only radiation pressure can explain the ULIG region of Figure \ref{figcie} (top right corner). Extending to larger SFR, our results indicate that radiation pressure will also be important for HLIGs (Hyperluminous Infra-red galaxies) \citep{row00}.
 Lastly, although it may appear that the role of radiation pressure in galaxies other
than ULIGs is less dominant than ram pressure as far as energetics is concerned, radiation pressure
may still play an important role in lifting the clouds to a large height before it is embedded in the
hot wind to help it survive long \citep{nath08, mur11}.
 
\section{Summary}
We have studied the outflows from disk galaxies driven by ram and radiation pressure and compared the theoretical results with
data in the parameter space of galaxy circular speed and SFR. We found that the driving mechanism of escaping wind is different
in low mass and high mass galaxies, with radiation pressure being important for
high mass galaxies with high SFR. Our results are also consistent with recently observed circumgalactic gas.

We thank Mitchell Begelman, Bruce Draine, Tim Heckman, Yuri Shchekinov and an anonymous referee for valuable comments.

\end{document}